# Detection Limits of NaI Scintillator Detector Based Aerial Source Detection Systems

Sebastian Ritter

*Ken and Mary Alice Lindquist Department of Nuclear Engineering*

*The Pennsylvania State University, University Park, PA 16802*

*Abstract*—Aerial source detection systems have the capability to rapidly provide radiological data over a large area of land [5][6][7]. Sodium Iodine (NaI) scintillator based aerial radiation detection systems of compact physical sizes have the potential aid nuclear security applications in a cost-effective manner when deployed on aerial vehicle systems. The Minimum Detectable Activity (MDA) of NaI scintillator airborne detectors is qualitatively evaluated as a function of detector-source distance and as a function of detector-source relative speed. It is found that the MDA increases exponentially with vehicle height and that MDA increases directly proportionally with the relative speed plus the square root of the relative speed. Furthermore, detection limits of an aerial detection system are evaluated in a case study. MDA is evaluated for the nuclear materials U-238 and Pu-239 as defined by the IAEA Safeguards Glossary Table II [12] and MDA is evaluated for arbitrarily selected isotopes found in Table I.2 of the IAEA document TECDOC-1344 [8].

## I. INTRODUCTION

Source detection methods in field instruments are typically made up of a three-step process of source detection, source location and source identification [4]. The relatively low spectral resolution of NaI aerial detector systems typically limit their applicability to source detection methods.

The scope of the problem faced by aerial source detection systems can be described as the determination of the presence of a radiation source via gamma source detection from an aerial detector. The Minimum Detectable Activity (MDA) is the lowest possible detectable activity of a source for a given detection environment, a given detection system and given confidence level. The source gamma energy spectrum may be altered by the source itself and its housing. As the gamma radiation travels through the environment it may alter its signature before it is detected by the NaI detector which in itself exhibits a certain response function further altering the detected gamma signature.

NaI Scintillator Detector Based Aerial Source Detection Systems comprise of a moving detector with a stationary source yielding short detection times with changing background counts. NaI scintillator detector based aerial source detection systems described in this paper are limited to passive detection systems with a constant background count rate.

## II. LITERATURE REVIEW

NaI Scintillator detector based aerial source detection systems may be employed by state governments to monitor border crossings. For example, the U.S. Department of Energy developed in cooperation with the U.S. National Nuclear Security Administration an Aerial Measuring System which constitutes a sodium iodine scintillator-based detector on both a fixed-wing airplane and a helicopter deployable system [6][7]. Schwarz et al. described [11] the application of an unmanned aerial vehicle mounted airborne gamma spectrometer system in Switzerland.

Aerial detection systems face several complications over ground based mobile detection systems. First, aerial detection systems are constrained by the size and weight limitations of an aerial platform. Second, the high speed of movement of fixed wing aerial detection systems limits the detection time of a fixed ground-based gamma or neutron source. Third, more rapid background fluctuations are expected due to geography changes when compared to ground based detection systems. Background count rates and fluctuations may also change with altitude and with weather. Forth, the larger source-detector distance leads to a larger field of view and a lower signal to noise ratio [9]. Steps in background count rate are expected as well as spectral changes in the gamma and neutron spectra due to increased gamma-ray and neutron scattering at large standoff distances [9].

One benefit of aerial detection systems is their ability of flying relatively closely to objects of interest due to their lack of reliance on ground infrastructure. Furthermore, their ability to reach large areas of land without reliance on ground infrastructure enables aerial detection systems to become a potential tool in border surveillance applications [10].

Detwiler et al. described [9] that land and water interfaces in addition to changing potassium, uranium and thorium background signals pose a challenge to quantitative measurements of source activity. The author describes that gamma down scatter dependability on altitude represents the most challenging issue for aerial source detection systems.

To the understanding of the author, a theoretical analysis on a NaI detection system's MDA performance as a function of speed and altitude has not yet been conducted for the corner case scenario of a detector window limited detection time.

### III. COMPONENTS

Cost of acquisition and operation of the aerial platform and the radiation detection system represent a significant aspect in design considerations of aerial detection systems. Design choices of detection systems may place requirements on aerial platforms and vice versa [6][7][11]. Survivability and stability of the detection systems, the distance between the source and the detector as well as the relative speed between the detector and the source may influence design choices. Chapter V B describes a specific aerial platform design within the framework of a case study.

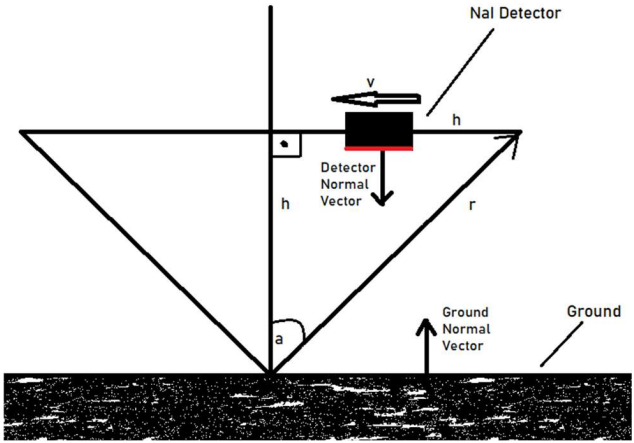

**Fig. 1.** Pictorial description of source and aerial platform geometries. The detector moves parallel to the ground at a relative speed of v and at a relative altitude of h. The detector has a 90-degree detection window and is pointed to the ground spawning a 45-45-90-degree triangle.

### IV. THEORY OF DETECTION

For low count rate detection systems, as typically found in aerial source detection systems, Currie et al. [1] first developed the most common definition of MDA as the minimum source activity required to make a binary determination with a certain confidence whether a source is present. Currie also arbitrarily chose a 0.05 false negative probability which yields a 0.95 detection probability. In non-laboratory environments, interference of the background leads to significant changes in the MDA. Gaussian counting statistics shall be valid and false negative error probabilities shall equal to 0.05. The MDA confidence level shall remain at a constant value of 95% for this paper.

The MDA is given by Eq. (1) [1].

$$MDA = \frac{N_D}{f\,\varepsilon_{abs}\,t} \qquad (1)$$

With $N_D$ being the minimum net value of source counts that meets the criteria of 5% false negative detection probability. The detection time for background counts and for gross source counts are set to be equal and denoted as t within the framework of this paper. f is the denotation for the radiation yield per disintegration of a given source and $\varepsilon_{abs}$ is the absolute detection efficiency. $\varepsilon_{int}$ is designated as the denotation for the intrinsic detector efficiency.

Background detection counts are assumed to be sequentially evaluated for source and background counts. The background measurement is assumed to be free of source radiation and the source strength is assumed to be weak compared to the background yielding a low signal to noise ratio. While the ratio of detection time of source counts over the detection of background counts can be optimized, this optimization is not subject to this study. For the detection time of source and background to be equal, $N_D$ can be expressed as Eq. (2) [1].

$$N_D = 4.65\sqrt{N} + 2.706 \qquad (2)$$

Whereas N is the number of counts recorded with only background. Since N is a function of detection time for a constant background rate, the two emerging aerial parameter dependencies are aerial speed and elevation as shown in Eq. (4).

Detector intrinsic efficiency is typically provided by detection system manufacturers. The relationship between the absolute detector efficiency $\varepsilon_{abs}$ and $\varepsilon_{int}$ is given as: $\varepsilon_{abs} = \varepsilon_{geo}\,\varepsilon_{int}$ where $\varepsilon_{geo}$ is a geometry depended factor which is further discussed in chapter IV C.

Within the scope of this paper, source hosing shielding is neglected. Source self-shielding is further neglected. Gamma radiation buildup in dry air is considered and all buildup radiation is assumed to be detected with the same intrinsic detector efficiency. Minimum Detectable Activity is evaluated for Gaussian counting statistics. The background detection count rate is assumed to be constant in time. The ground-based source emission rate is assumed to be constant in time. The background detection count rate is further assumed to be constant in vehicle speed and altitude. All buildup radiation is assumed to be detected with the same intrinsic detector efficiency as the original gamma source energy. The detector signal processing is assumed to be converting 100% of all intrinsic scintillation events into recorded electrical signals. Any environmental ground backscattering is neglected. These simplifications may underestimate MDA significantly. However, one may learn about the qualitative MDA behavior as a function of vehicle speed and altitude.

*A. Source Activities of Interest*

Dangerous Source Activities of selected radionuclides are provided by Table I.2 in TECDOC-1344 [8]. In this paper, MDA for the isotopes Cs-137, I-131 and Co-60 are evaluated.

The most prominent radionuclide gamma signatures can be read from the IAEA Nuclear Data Services with data sourced from ENSDF. All values are written here are approximate and as provided by the IAEA database. Cs-137 features a prominent 662 keV line with a decay yield of 80.6%. Co-60 exhibits a prominent 2506 keV line at a decay yield of 99.9%. I-131 exhibits a prominent 365 keV line with a decay yield of 73%.

IAEA significant quantities of nuclear material can be read from the IAEA Safeguards Glossary Table II [12]. Significant quantities are 8 kg plutonium for bulk plutonium containing less than 80% Pu-238, 8 kg of U-233, 25 kg of U-235 for uranium enriched to 20% or more, 75 kg of U-235 for uranium enriched to less than 20% and 20 metric tons of thorium. Thorium is not considered in this paper due to the large amount of material required to be considered of significant quantity [12][14]. 20 metric tons of thorium may be detected by other methods.

For highly enriched uranium, the principal gamma ray lines have energies of 186 keV, 1001 keV and 2615 keV [13]. 186 keV gammas may be shielded easily. The 1001 keV line is uniquely associated with the presence of U-238 and may be an indicator for the presence of HEU. The 2615 keV is roughly two orders of magnitude weaker than the 1001 keV line [13], making the 1001 keV line the most interesting feature of uranium for aerial detection. Pu-239 may be detectible by a prominent 414 keV line.

Estimating the radiation yield of nuclear material is done for Pu-239 and for U-238 by dividing the specific activities of selected gamma ray lines by the source specific activity leading to the gamma yield in units of gammas per decay. A brief calculation leads to the values of $5.9*10^{-3}$ for the 1001 keV U-238 line and $1.5*10^{-5}$ for the 414 keV Pu-239 line.

The source specific activities are taken from the CRC handbook [21] and are read as $2.29*10^9$ Bq/g for Pu-239 and 12 400 Bq/g for U-238. The specific activities of the 414 keV Pu-239 gamma ray line is obtained from Table 5 of Strohmeyer's thesis [19] as $3.43*10^4$ gammas per gram per second. The specific activities of the 1001 keV U-238 line is obtained from Reilley's Table 2-1 [20] as 73.4 gammas per gram per second.

### B. Minimum Count Net Value $N_D$ Evaluation

In order to evaluate the detection time for a given source, an arbitrary model of the source, environment and detection system is set up.

The aerial vehicle is travelling at a constant speed and altitude relative to the ground. A source of gamma radiation shall be in the detectable window of the detector if the angle between the line of sight of the source and vector at the location of the detector, which is pointing to the ground, is less than or equal to 45 degrees. This criterion is shown in Fig. 1 and spawns a 45-45-90-degree triangle. It is evident that the detector in this configuration has a detection window of approx. 1.84 sr.

Using Trigonometric ratios, the detection time can be calculated as a function of altitude and ground speed in Eq.3.

$$t = 2h/v \qquad (3)$$

Whereas t is the detection time, h is the aerial system height above ground and v is the speed relative to the ground. $c_B$ represents the constant background count rate. Inserting Eq. (3) into Eq. (2) yields Eq. (4).

$$N_D(h,v) = 4.65\sqrt{\frac{c_B 2h}{v}} + 2.706 \qquad (4)$$

### C. Geometry Factor $\varepsilon_{geo}$ Evaluation

The source is assumed to be unshielded with a direct line of sight to the detector when within the 90-degree wide detection window. Only gamma rays are assumed to be detectable and all neutron induced gamma emissions are neglected.

The source gamma photons interact with their environment via photoelectric absorption, Compton scattering and pair production for incident gammas above the 1022 keV energy threshold. Fritz S. et al. computed the exposure buildup factors of air [3]. The authors considered gamma coherent scattering, bremsstrahlung, bound-electron Compton scattering, annihilation gamma rays and K-L-Edge X-rays using the XCOM photon cross section library.

The buildup flux at the detector is given by Eq. (5) [15].

$$\phi_b(r,E) = \frac{S(E)\, B(\mu(E)r)e^{-\mu(E)r}}{4\pi r^2} \qquad (5)$$

The energy dependent gamma attenuation coefficient in air is given by $\mu(E)$. $B(\mu(E)r)$ is the buildup factor that is dependent on the factor $\mu(E)*r$. The buildup factor $B(\mu(E)r)$ must approach 1 for r = 0 since there is no buildup if no scattering can occur. The factor $\mu(E)*r$ equals the detector-source distance r in units of mean free path (mfp) and the expression $1/\mu(E)$ yields the mfp. $\phi_b(r,E)$ is the neutron flux in units of gammas per square centimeter per second. S is the gamma emission rate in units of gammas per second. For a detector area A, whose normal vector is facing parallel the ground normal vector as seen in Fig. 1, the incident gamma rate is given by the factor $\phi_b(r,E)*A$. The number of pulses recorded by the detector per second is given by $\phi_b(r,E)*A*\varepsilon_{int}(E)$. Multiplying this expression by the counting time and inserting Eq. (3) for time t yields the number of expected counts in Eq. (6).

$$N(r,E,v,h) = \phi_b(r,E)*A*\varepsilon_{int}(E)*2h/v \qquad (6)$$

To remove the distance r dependence in Eq. (6), one must integrate over all distances r. The maximum distance r is $\sqrt{2}h$ and the minimum is h, making these values the integral limits. Note that $\phi_b(r,E)$ is symmetrical around r = H introducing a factor 2 in Eq. (8).

The approximation $r \rightarrow \bar{r}$ is made to allow for simpler integration. The detector source distance r is being approximated by the average detector source distance in Eq. (7).

$$r \to \bar{r} = \frac{\sqrt{2}h + h}{2} = 1.207\, h = \bar{r}(h) \quad (7)$$

It follows that the buildup factor $B(\mu(E)r)$ in Eq. (5) is evaluated at $B(\mu(E)\bar{r})$. Integration of Eq. (6) over r is shown in Eq. (8) to Eq. (10).

$$N(E,v,h) \equiv \int_h^{\sqrt{2}h} N(\bar{r}, E, v, h)\, dr \quad (8)$$

$$N(E,v,h) = \phi_b(1.207\,h, E) * 2h(\sqrt{2}-1) * A * \varepsilon_{int}(E) * \frac{2h}{v} \quad (9)$$

$$N(E,v,h) =$$
$$= \phi_b(1.207\,h, E) * 2h(\sqrt{2}-1) * A * \varepsilon_{int}(E) * \frac{2h}{v}$$
$$= \frac{B(\mu(E)1.207\,h) e^{-\mu(E)1.207\,h} S(E) * A * \varepsilon_{int}(E) * 0.0905}{v} \quad (10)$$

Eq. (10) is only valid for h > 0 and v > 0. Due to the aerial detection system geometry in Fig. 1, the buildup factor introduces the sole height dependency in Eq. (10).

The simplification $r \to \bar{r}$ introduces an energy dependent maximum absolute error in the Eq. (10). Chapter V A paragraph two explains that the buildup function can be described as a positive monotone function using buildup data obtained by Fritz S. et al. [3]. The neutron flux $\phi_b(r, E)$ is a linear function in the buildup $B(\mu(E)r)$, decreases with $1/r^2$ and decreases exponentially with $e^{-\mu(E)r}$. The buildup function itself increases monotonically with distance. However, it's the weak third and second power dependency on the distance r, as seen in Eq. (15), is dominated by the exponential decline $e^{-\mu(E)r}$ at distances far from the radiation source. One concludes that $N(r, E, v, h)$ decreases monotonically with increasing distance at large distances.

The positive monotonic nature in r of the function $N(r, E, v, h)$ allows for a simple estimation for the maximum absolute error in Eq. (10) introduced with the simplification $r \to \bar{r}$. The energy dependent maximum absolute error in the number of expected counts $N_{abs\,err}(E, v, h)$ is given in Eq. (11).

$$N_{abs\,err}(E,v,h) = \text{MAX}\left( \left| \int_h^{\sqrt{2}h} N(r=\sqrt{2}h, E, v, h)\, dr - \int_h^{\sqrt{2}h} N(r=\bar{r}, E, v, h)\, dr \right|, \left| \int_h^{\sqrt{2}h} N(r=h, E, v, h)\, dr - \int_h^{\sqrt{2}h} N(r=\bar{r}, E, v, h)\, dr \right| \right) \quad (11)$$

The absolute detection efficiency in Eq. (1) must be determined to obtain the MDA of a radioactive sample. The absolute and intrinsic efficiencies are connected via a geometry factor. The geometry factor $\varepsilon_{geo}$ can be calculated as follows.

$$\varepsilon_{abs} \equiv \frac{\text{number of counts produced by the detector}}{\text{number of gammas emmitted by the source}} \quad (12)$$

$$\varepsilon_{abs} \equiv \frac{N(E,v,h)\frac{\varepsilon_{int}}{\varepsilon_{int}}}{S(E) * 2h/v} = \varepsilon_{geo}\, \varepsilon_{int} \quad (13)$$

$$\varepsilon_{geo}(E,v,h)$$
$$= \frac{B(\mu(E)1.207\,h) e^{-\mu(E)1.207\,h} * A * 0.04525}{h} \quad (14)$$

The MDA can now be determined by inserting Eq. (4) and Eq. (14) into Eq. (1).

$$MDA(E,v,h) = \frac{N_D}{f\, \varepsilon_{abs}\, t} =$$
$$= \frac{v * \left(4.65 \sqrt{\frac{c_B 2h}{v}} + 2.706\right)}{f\, \varepsilon_{int}\, B(\mu(E)1.207\,h) e^{-\mu(E)1.207\,h} * A * 0.0905} \quad (15)$$

An absolute MDA error is introduced by the simplification $r \to \bar{r}(h)$. The magnitude of this absolute error may be estimated with Eq. (11). Using Eq. (11) and Eq. (15) one obtains Eq. (16).

$$MDA_{abs\,err}(E,v,h) = \text{MAX}(|MDA(r=\sqrt{2}h, E, v, h) - MDA(r=\bar{r}, E, v, h)|,\ |MDA(r=h, E, v, h) - MDA(r=\bar{r}, E, v, h)|) \quad (16)$$

A statistical error in the MDA is introduced by the factor $N(E,v,h)$ in the expression $\varepsilon_{abs}$ found in Eq. (13). The number of expected detector counts $N(E,v,h)$ is subjected to errors due to the statistical nature of radiation detection. For Gaussian statistics, the expected error is given by the expression $\sqrt{N(E,v,h)}$. The statistical error in MDA is evaluated at the position $r = \bar{r}$ for a given E, v and h, which may be denoted as $MDA(r=\bar{r}, E, v, h)$.

Following the $r \to \bar{r}$ absolute error evaluation in Eq. (16), error propagation is applied to Eq. (15) to compute the final error which comprises of the sum of the absolute approximation error and the one sigma counting statistics error. To determine the factor $\sqrt{N(E,v,h)}$, Eq. (10) is utilized while subsidizing the factor S(E) with the center value of the calculated MDA.

*D. Background Signal Evaluation*

Bruce L. Dickson et al. [16] reviewed background noise reduction methods used in potassium, uranium and thorium ground concentration aerial mapping surveys. The author identifies high frequency variations in detector count rates due to potassium, uranium and thorium concentrations in the ground, airborne radon, variable vegetation and soil moistures, energy drifts in radiation detectors, aircraft movements and cosmic radiation [16]. Background radiation may vary significantly in time [2]. For example, airborne radon has been shown to change with precipitation [2][17]. Background radiation also varies by altitude and solar activity [2][17][18].

V. METHOD OF DEPLOYMENT

Aerial source detection systems may be deployed on the southern border of the United States for the detection of nuclear and radiological smuggling operations. The MDA performance of such systems is evaluated theoretically and in a specific case study.

*A. Theoretcal MDA Performance Validation*

Eq. (15) provides a gamma ray energy dependent method for MDA estimation. For source spectra of a single energy $E_0$ with a strength of $S_0$ gammas per second, S(E) can be written as the delta function S(E) = $S_0 \delta(E - E_0)$. For certain gamma sources the uncollided gamma ray energy spectrum at large distances can be approximated as a single delta function. The large source distance approximation enables the negligence of source x-rays and low energy gamma ray detection, since these can be assumed to be completely absorbed before reaching the detector. The radioisotopes Cs-137, I-131 and Co-60 each feature one high energy gamma line with a greater than 70% radiation yield. For these isotopes the delta function approximation is made and Eq. (15) may be integrated over all source gamma ray energies yielding a height h and speed v depended $MDA\ (E = E_0, v, h)$.

Gamma rays do not have a well-defined range due to the exponential decline of their uncollided beam which is given by $I = I_0 * e^{-\mu x}$. The mean free path (mfp) of gamma rays is defined as the distance when the intensity of a parallel beam of gamma falls to 1/e of its initial value. The mfp $\lambda$ for gamma rays in air can be calculated as $\lambda = 1/\mu$ [15] with the gamma attenuation coefficient $\mu$. Using Appendix 2.4 in Lamarsh Baratta [15] one obtains a mfp of 51m for 100 keV gamma rays in dry air at 1 atm and at 0 °C. The mfp for 600 keV gammas in dry air is calculated as 97 m.

For source-detector distances of greater than 100 m, the intensity of 100 keV or less gamma rays has decreased by two mean free paths or 1/e² and is considered negligible. Thus, only buildup factors for gamma energies of greater than 100 keV are of interest for aerial detection systems with a source-detector distance of 100 m or greater. This approximation allows for the negligence of source X-Rays whose energy spectrum typically peaks in intensity at energies below 100keV.

Exposure buildup factors in air by Fritz S. et al. [3] increase monotonically in gamma energy form 10 keV to 100 keV and follow a monotone decrease from 100 keV to 10 MeV. For a given gamma energy E, the buildup factors increase monotonically with increasing distance.

For a given distance r, a relationship between exposure buildup factors in air and the gamma energy is determined to provide for a functional expression of the buildup factor as a function of energy E. For energies greater than 100 keV and for distances greater than 1 mfp in dry air, buildup factors provided by Fritz S. et al. [3] fit potential functions with a R² factor of 0.96 for 100 keV and a R² factor of greater than 0.98 for 1 to 10 MeV gamma ray radiation. Five of these energy depended buildup functions are determined at source distances of 1, 2, 5, 10 and 20 mfp.

One can now tabulate the buildup factors for several energies for each of these five functions. This yields five data sets, each valid at a certain distance r, listing the buildup for a given gamma ray energy. For a constant gamma energy, a function of the distance dependent buildup may now be fitted to these entries. The buildup, as a function of distance r in units of mean free path, is found to fit a polynomial of third power with an R² value of greater than R² = 0.9999 for gamma energies between 10 keV and 10 MeV. This process is repeated for each gamma ray energy of interest, yielding different buildup functions each for a selected gamma ray energy. MDA values in table 1 and table 2 were obtained following this process.

The energy and height dependent buildup factor $B(\mu(E) 1.207\ h)$ in Eq. (15) can now be approximated as a function that is solely dependent on the parameter h. In this chapter, the gamma ray energy evaluated is arbitrarily chosen to be 600 keV. The attenuation coefficient for 600 keV gammas is determined as $\mu(600\ keV) = 0.0104\ (1/m)$ [15]. The buildup function can be written as Eq. (17).

$$B(\mu(600\ keV)\ r) = 10^{-8} r^3 + 3 * 10^{-5} r^2 + 0.0203 r + 0.2176 \qquad (17)$$

Where r is the detector-source distance. For the approximation $r \to \bar{r} = 1.207\ h$ we obtain Eq. (18).

$$B(\mu(600\ keV)\ 1.207\ h) \equiv B(h) = 1.76 * 10^{-8} h^3 + 4.37 * 10^{-5} h^2 + 0.0245 h + 0.2176 \qquad (18)$$

Inserting Eq. (18) into Eq. (15) we obtain the MDA of 600 keV gamma rays in Eq. (19).

$$MDA\ (E = 600\ keV, v, h) = \frac{v * (4.65 \sqrt{\frac{c_B 2h}{v}} + 2.706)}{f\ \varepsilon_{int}\ B(h) e^{-\mu(E) 1.207\ h} * A * 0.0905} \qquad (19)$$

It is evident that the MDA increases exponentially with height h and increases with the sum of the square root of the speed v plus a linear function v. The background count rate $c_B$ contributes as a square root function to the MDA.

Eq. (19) is plotted in Fig. 1 for a constant background count of 20 cps, a radiation decay yield of 0.806, a detector area of 400 cm² and an intrinsic detector efficiency of 15%.

The 600 keV gamma line is used as a placeholder for the 662 keV Cs-137 gamma line. At a speed of 37 m/s and height of 100 m the Cs-137 MDA is calculated as 5.9 MBq with an uncertainty of 4.2 MBq. At 1000 m the Cs-137 MDA is 43.3 GBq with an uncertainty of 30.8 MBq. In comparison, at a speed of 370 m/s and height of 100m the MDA is calculated as 20.9 MBq with an uncertainty of 14.9 MBq. The MDA error is calculated using methods described in the last paragraph of chapter IV C. Fig. 2 illustrates the MDA dependency on detector relative speed v and height h. Increasing detector height h exponentially increases MDA whereas an increase in speed v impacts the MDA to a lesser degree. As the speed is reduced to 0 m/s the detection time approaches infinity and the MDA approaches 0 Bq.

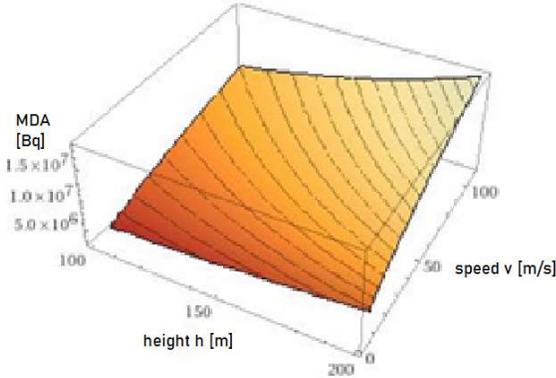

**Fig. 2.** Plot of the MDA against aerial vehicle speed v and altitude h. Darker orange symbolizes a lower MDA. The black lines represent regions of constant MDA. For a height difference of 100m, the MDA changes by one order of magnitude.

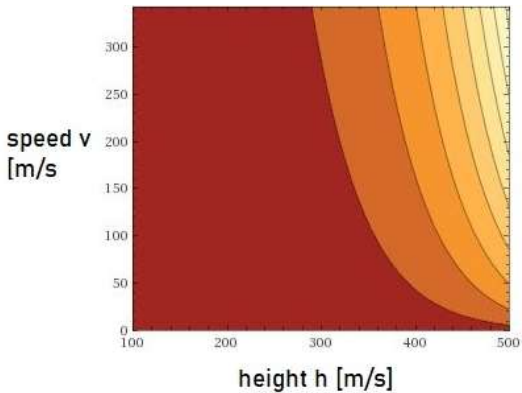

**Fig. 3.** Contour lines plot of the MDA against aerial vehicle speed v and altitude h for the radioisotope Cs-137. Darker orange symbolizes a lower MDA. As the speed approaches zero, the MDA approaches zero, representing infinite counting time. The height is plotted from 100m because the buildup factor approximation in Eq. (17) is only valid for heights h greater than one mean free path. For Cs-137 1 mfp equals roughly 100m.

*B. Case Study*

Detection limits are further determined for a case study. An aerial detection system is mounted on a 2019 Cessna Skyhawk [23] fixed wing airplane. The plane is operating at a GPS controlled altitude of 123 meters or 400 ft above ground. The plane's air speed is selected at 37 m/s, 150% of the plane's stall air speed. The windspeed relative to the ground is assumed to be 0 m/s. Full fuel payload is 254 kg. With a pilot and co-pilot mass totaling 200 kg, the detection system is constrained to a mass of no more than 54 kg. 50% of the mass of the detector system is reserved for environmental isolation for the NaI detectors to increase recalibration frequency and reduce damage induced by vibration, temperature and humidity changes. This yields a detection system mass of only 27 kg. This mass requirement is fulfilled by the "SAMmobile 150 (RD-150)" gamma detection system with a total mass of 20 kg [22]. The NaI scintillator of this system has a size of 2x4x16 inches yielding a detector window of roughly 400 cm². The co-pilot uses a tablet PC for data evaluation and is tasked with alerting the pilot in the case of positive source detection. Detector intrinsic efficiency is assumed to be 15%. A constant background of 20 cps is again assumed. Note that the mean free path of 1 MeV gammas for dry air at IUPAC standard temperature and pressure is 123.3 m.

Table 1 lists the MDA for the three IAEA radiological materials of interest under conditions of this case. The last column in Table 1 denotes whether the radioisotopes of interest can be detected to activities designated as dangerous. The confidence level for the MDA is 95%.

Table 2 shows the MDA for IAEA nuclear materials of interest under conditions of this case. The last column in Table 2 denotes whether IAEA significant quantities of nuclear materials can be detected. U-238 is assumed to be the only indicator for the presence of IAEA significant amount of nuclear material for the category of 75 kg of U-235 for uranium enriched to less than 20%. For this category, a minimum of 375 kg of U-238 would need to be observed for positive detection. The statistical error introduced in the MDA by the factor $N(E, v, h)$ is evaluated to be at least one order of magnitude smaller than the absolute error introduced by the $r \to \bar{r}$ approximation and is found to be negligible for the final MDA uncertainty values in Table 1 and Table 2. The confidence level for the MDA is again 95%.

MDA in Table 1 and Table 2 is calculated using Eq. (15). Absolute uncertainties are calculated using Eq. (16). Statistical counting errors are then propagated through Eq. (15) and added to the absolute uncertainties introduced by Eq. (16). Simplifications made in the MDA evaluations in Table 1 and Table 2 are described in the last paragraph of the introductory section of Chapter IV. These strong simplifications lead to a severely underestimated MDA.

TABLE I
CALCULATED MDA OF SELECT RADIOISOTOPES

| Substance | MDA (Bq) | MDA Uncertainty (Bq) | IAEA Visible ? |
|---|---|---|---|
| Co-60 | 6.9*10^6 | 5.0*10^6 | yes |
| I-131 | 6.1*10^6 | 4.4*10^6 | yes |
| Cs-137 | 5.9*10^6 | 4.2*10^6 | yes |

TABLE II
CALCULATED MDA OF SELECT IAEA NUCLEAR MATERIALS

| Source | MDA (Bq) | MDA Uncertainty (Bq) | MDA (kg) | MDA Uncertainty (kg) | IAEA Visible ? |
|---|---|---|---|---|---|
| U-238 | 9.0*10^8 | 6.5*10^8 | 72.6 | 52.4 | no |
| Pu-239 | 2.8*10^11 | 2.0*10^11 | 0.12 | 0.09 | yes |

## VI. Conclusion

A theoretical analysis on a NaI detection system's MDA performance as a function of speed and altitude has been conducted for the corner case scenario of a detector window limited detection time.

Under simplified environmental and detection system conditions, it is shown that the MDA increases exponentially with relative height h. MDA also increases as a function of the square root of the relative speed plus a linear function of the relative speed. The background count rate $c_B$ contributes as a square root function to the MDA. To maximize MDA, the reduction of the source-detector distance is to be prioritized over a reduction in detector-source relative speed.

A case study was found to imply that a passive aerial detection system may be suitable for radioisotope source detection and may not be suitable for nuclear material detection.